\documentclass[10pt]{wlscirep}
\usepackage[utf8]{inputenc}
\usepackage[T1]{fontenc}
\usepackage{booktabs}
\usepackage{url}

\DeclareUnicodeCharacter{2212}{-}

\graphicspath{{/}}

\title{\textit{Inter-Body Coupling} in Electro-Quasistatic Human Body Communication: Theory and Analysis of Security and Interference Properties}

\author[1,*]{Mayukh Nath}
\author[1]{Shovan Maity}
\author[1]{Shitij Avlani}
\author[2]{Scott Weigand}
\author[1]{Shreyas Sen}
\affil[1]{School of Electrical and Computer Engineering, Purdue University}
\affil[2]{Eli Lilly and Company}

\affil[*]{nathm@purdue.edu}

\begin{abstract}
Radiative communication using electromagnetic fields is the backbone of today's wirelessly connected world, which implies that the physical signals are available for malicious interceptors to snoop within a 5-10 m distance, also increasing interference and reducing channel capacity. Recently, Electro-quasistatic Human Body Communication (EQS-HBC) was demonstrated which utilizes the human body's conductive properties to communicate without radiating the signals outside the body. Previous experiments showed that an attack with an antenna is unsuccessful, more than 1 cm of the body surface and 15 cm of an EQS-HBC device. However, since this is a new communication modality, it calls for investigation of new attack modalities - that can potentially exploit the physics utilized in the EQS-HBC to break the system. In this study, we present a novel attack method for EQS-HBC devices, using the body of the attacker itself as a coupling surface and capacitive inter-body coupling between the user and the attacker. We develop theoretical understanding backed by experimental results for inter-body coupling, as a function of distance between the subjects. We utilize this newly developed understanding to design EQS-HBC transmitters to minimize the attack distance through inter-body coupling as well as minimize the interference among multiple EQS-HBC users due to inter-body coupling. This understanding allows us to develop more secure and robust EQS-HBC based body area networks in the future.

\end{abstract}
\begin{document}

\flushbottom
\maketitle
%
%
\thispagestyle{empty}

\section*{Introduction}

\textit{Wireless communication using electromagnetic radiation} has formed the base-bone for today's ubiquitous connected devices with a possibility of trillions of connected `things' forming the `Internet of Things' (IoT) revolution. A portion of these IoT devices will be on, around or even inside the human body creating a network of intelligent devices - namely the `Internet of Body' (IoB).  The distinguishing feature for IoB devices compared to IoT devices is that IoB devices share a common medium - i.e. the body itself\cite{sen_tedx2019}.

\begin{figure}
    \centering
    \includegraphics[width=\textwidth]{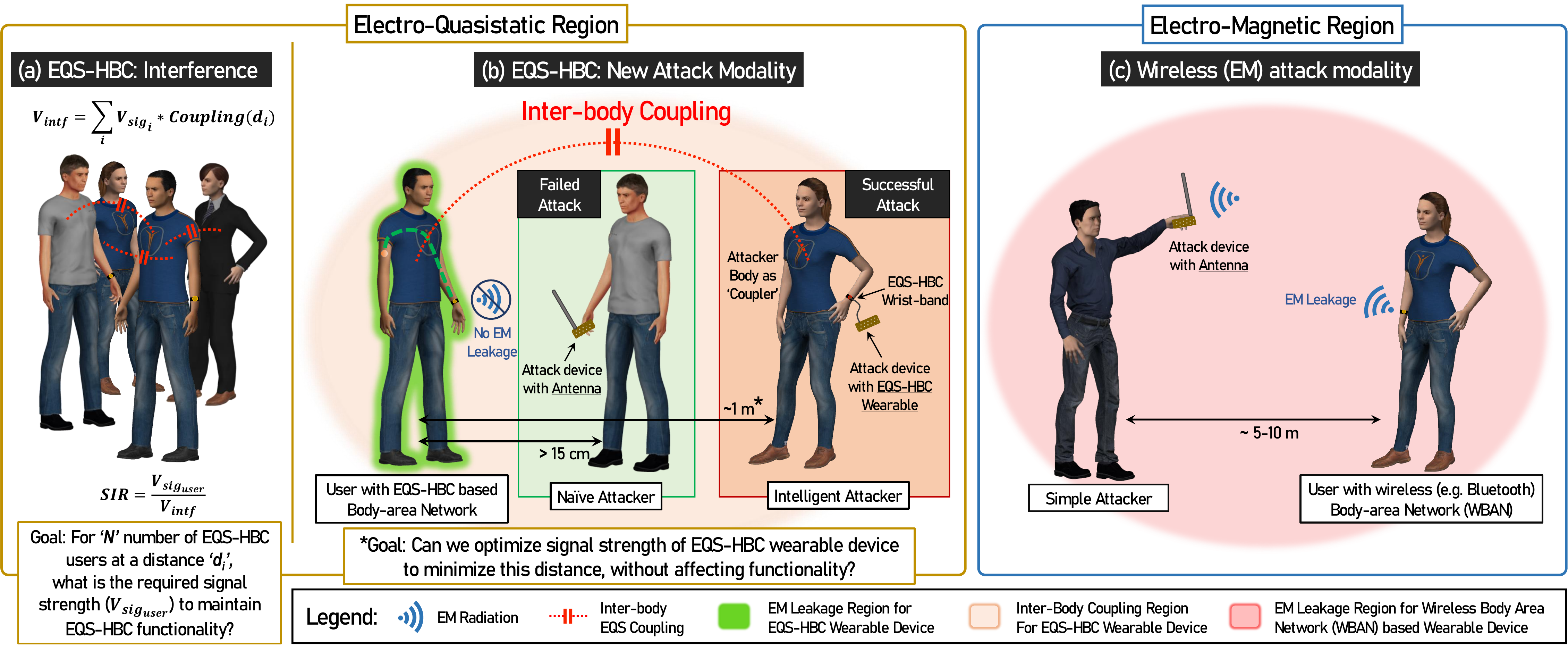}
    \caption{Inter-body coupling in Electro-quasistatic region: (a) Interference in received EQS-HBC signal due to inter-body coupling with other users. For multiple EQS-HBC users in close proximity, the received signal is usable, only if the interference signal is a few dB lower than the signal. (b) While EQS-HBC devices restrict EM leakage within 10 cm of the user's body, inter-body capacitive coupling can give rise to a new attack modality, where the attacker's body is used to capacitively couple to the user's body, and the coupled signal is picked up using an EQS-HBC receiver. (c) For devices that do not restrict EM leakage, such as Bluetooth or other WBAN devices, the signal can be picked up be an attacking device with an antenna within 5-10 m of the user. \textit{The
human figures were created using the open-source software `MakeHuman'\cite{makehuman}.}} \label{fig:introduction}
\end{figure}

Since in traditional Body Area Network (BAN) devices operate through radiative communication such as Bluetooth, Med-Radio, WiFi etc, the physical signals are not only available on and around the user's body, but also broadcast away from the user, making it available for malicious interceptors within 5-10 m distance (figure \ref{fig:introduction}c). This brings us to the natural question that can the distinguished feature, i.e. the body as a common medium, can be used to improve the security of IoB devices?

Recently, Electro-Quasistatic Human Body Communication (EQS-HBC)\cite{nsr_das_19} was introduced as a "Physically Secure" way to communicate among IoB devices using the body itself as a `wire' \cite{bodywire_jssc_19}. Unlike traditional WBAN devices, frequencies used in EQS-HBC are low (< 1 MHz) - such that the corresponding wavelength is large with respect to the human body, making the communication electro-quasistatic (EQS) in nature. EQS-HBC, more specifically capacitive EQS-HBC, uses the human body as a forward path in a circuit to transmit signal between a transmitter and receiver, and completes the communication path through parasitic capacitive coupling formed between the EQS-HBC device's floating ground and earth's ground. As the human body, acting as electrically small antenna, does not radiate well in EQS frequency regime, it makes the EQS-HBC communication path analogous to a closed loop electrical circuit. As demonstrated by Das et al\cite{nsr_das_19}, since far-field electromagnetic radiation of signal is prevented in EQS-HBC signal, the signal is restricted within 1 cm of the body surface and 15 cm of an EQS-HBC device - thus making it physically secure. 


 However, one may raise a question, whether an E-field probe or an antenna is indeed the best way to attack or sniff EQS-HBC communication. We motivate this discussion by considering the fact that EQS-HBC uses the human body as medium, and asking the question: Is there any way two human bodies can couple, making EQS-HBC signals available on a second person's body? A literature survey reveals that there has been studies that have considered the human body as an antenna before, and these works fall under mainly two categories - one where the Specific Absorption Rate (SAR) of the human body has been investigated over different frequencies \cite{tomovski_sar_2011,kibret_sar_2014,kibret_cyl_2015}, and the other, where the interference received by the human body for incident EM waves has been examined \cite{kibret_antenna_hbc_2014, Hwang_interf_2017}. Kibret\cite{kibret_antenna_hbc_2014, kibret_monopole_2015} characterized antenna properties of the human body by modelling it as a monopole antenna in the 1-200 MHz range. Li\cite{Li_antenna_2017} used the same approach to examine wireless signal transmission between two humans for frequencies 1-90 MHz. The frequency range explored in these works fall out of the low frequency EQS-HBC range (< 1 MHz) however - and a \textit{theory of human inter-body coupling}, to the author's best knowledge, has never been developed before. In this paper, we answer the question of a better attack modality of EQS-HBC by developing, for the first time, an in-depth understanding of inter-body coupling over a broad frequency range (100 kHz - 1 GHz), and showing that the human body can function as a capacitor plate in the EQS region and an attack device connected to the attackers body can potentially "sniff" EQS-HBC signals from a further distance, compared to an attack device connected to an antenna (figure \ref{fig:introduction}b). Using the developed theory and understandings of the physical principles, we propose an improved EQS-HBC communication design that is tolerant of "Inter-Body Attack" as well as minimizes inter-human interference, improving channel capacity.
 



\begin{figure}[!ht]
    \centering
    \includegraphics[width=0.9\textwidth]{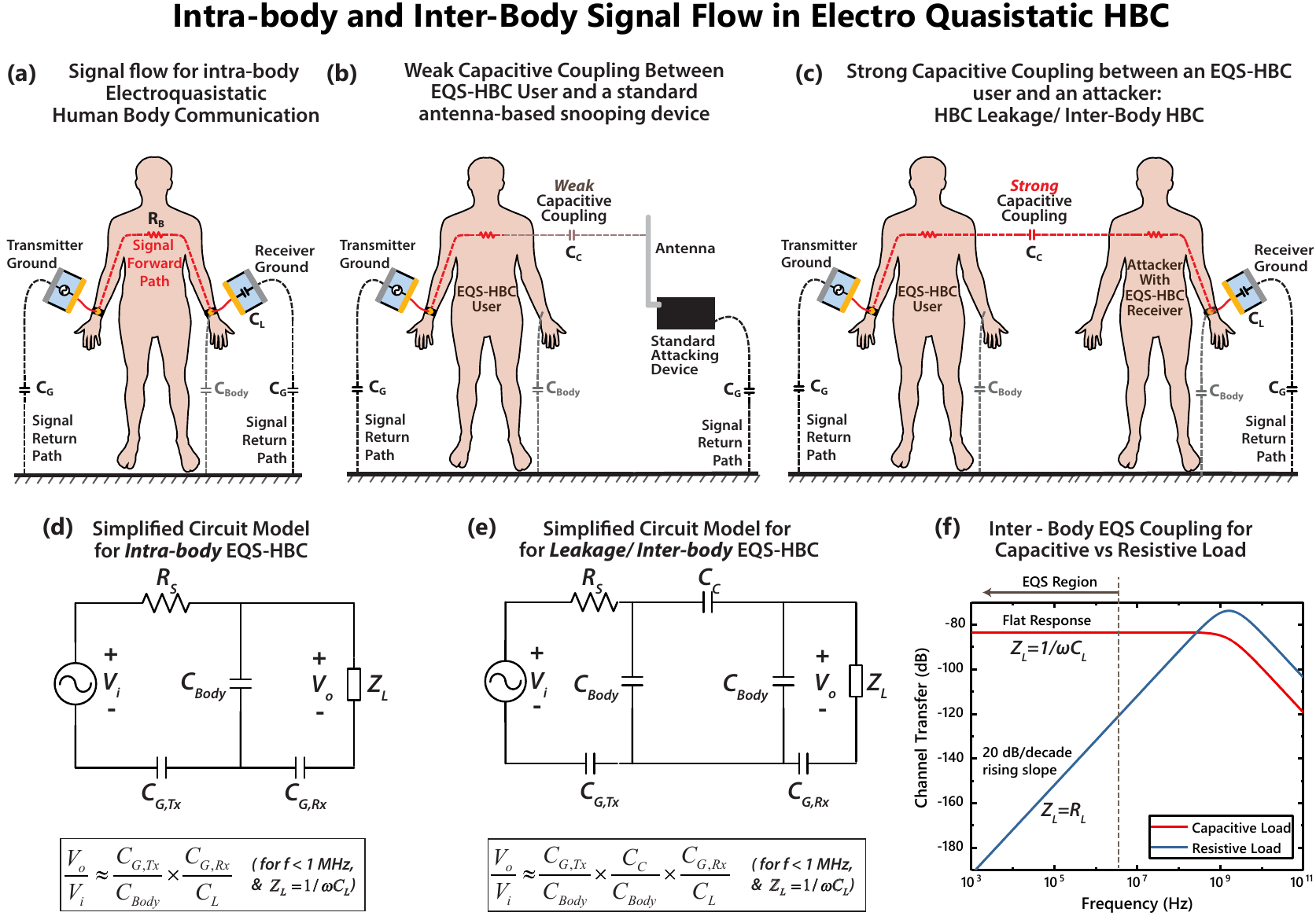}
    \caption{(a) Forward and return path for regular intra-body Electro-quasistatic Human Body communication (EQS-HBC). Forward path is formed through the human body, while the return path is formed through parasitic capacitances $C_{G,Tx}$ and $C_{G,Rx}$ with environment. (b) Weak Capacitive coupling between an EQS-HBC user and an antenna, ensuring minimal leakage pick-up by that antenna. This ensures minimal interference and maximum security towards antenna based devices. (c) Strong capacitive coupling, $C_C$ between two human bodies present the question of inter-body signal leakage for EQS HBC. This can potentially allow the 2nd user, the attacker, to sniff EQS HBC signals from the 1st user. If the 2nd person is just a regular user of HBC, the capacitive coupling can cause interference between the EQS HBC Signals from the two bodies. (d) Simplified circuit model for regular intra-body HBC of figure \ref{fig:concepts}a, and approximate expression for channel loss. (e)Simplified circuit model for inter-body HBC or HBC leakage, from figure \ref{fig:concepts}b, and approximate expression for channel loss. The extra term $C_C/C_{Body}$ causes an additional loss for inter-body coupling. (f) Comparison of inter-body coupling for capacitive vs Resistive load at the receiver's end. For capacitive load, the low-frequency region is a flat-band response. For resistive load, the response is a 20 dB/decade rising slope.} \label{fig:concepts}
\end{figure}

\subsection*{EQS Inter-Body Coupling}

In capacitive EQS-HBC, Signal electrodes of the transmitting and receiving devices are connected to a human body, while the ground electrodes are left floating. As shown in figure \ref{fig:concepts}a, human body forms the forward path of communication\cite{maity_forward_loss_2018}, while the return path is formed by parasitic capacitance between the earth's ground and the transmitter and receiver ground planes\cite{nath_return_cap} ($C_{G,Tx}$ and $C_{G,Rx}$ respectively). This parasitic return path is key in capacitive EQS-HBC operation, as low-frequency EQS operation makes the system analogous to an electrical circuit(\ref{fig:concepts}d) - where a closed loop must be present between the transmitter and receiver. The impedance of the return path capacitances are much higher compared to the forward path resistance $R_{B}$ for frequencies $<$ 1 MHz\cite{nath_return_cap, bio_physical_tbme_19}, and when operated in that region, most of the potential drop happens across $C_{G,Tx}$ and $C_{G,Rx}$. The fact that wavelenth of signals are much larger than hunan body size, leaves the human body roughly at the same quasistatic electric potential, letting us to incorporate the body as a point node in the circuit model and introduce the idea of inter-body coupling simply in terms of a lumped version of the distributed coupling capacitance $C_C$, as shown in figure \ref{fig:concepts}c. As the primary EQS-HBC user's body stays at a constant electrostatic potential at a given point in time, this inter-body capacitance $C_C$ can couple part of that potential to a second person's body, and can potentially be picked up by an EQS-HBC device on his body. This inter-body coupling affects EQS-HBC in two different ways, namely security and interference:

\subsubsection*{Security Perspective: New Attack Modality with the Human Body as a Capacitor Plate}

Physical security of EQS-HBC has been demonstrated\cite{nsr_das_19} using E-field probes or standard RF antennas to pick up signal leakage from an HBC user. However, these probes and antennas are inefficient at the low frequency range of EQS HBC. For example, for an operating frequency of 100 kHz, an efficient mono-pole antenna will have to have a length of 750 meters, which is completely impractical. However, in these low frequency range, these 'antennas' can also pick up signal by capacitively coupling to the body of an EQS-HBC user. Now, typical electrical antennas tend to have a very small surface area, thus forming an inefficient capacitive coupling. Ideally, an electrode with a huge surface area should be able form a much better capacitive coupling with the body of an EQS-HBC user, and the \textit{easiest movable semi-floating large surface area} available to an attacker is his or her own body itself. Figure \ref{fig:introduction}b, illustrates a probable attack scenario where a naive attacker with an antenna placed more than 15 cm away from an EQS-HBC user is unable to snoop the signal, whereas an informed attacker with an EQS-HBC receiver successfully does the same by using her body as a capacitive coupler and staying at a longer distance - as long as the coupling is strong enough to provide enough signal at the snooping device, and thus potentially breaking the physical security of EQS-HBC using this novel 'Inter-Body Attack'.

\subsubsection*{Interference Perspective: Proximity between multiple EQS-HBC users and impact on SIR}

Inter-body capacitive coupling for EQS-HBC also poses another potential problem - that of interference - between multiple HBC users in close proximity, where the signal from one user's body can interfere with that on the other user's body. As illustrated in figure \ref{fig:introduction}a, for $N$ number of additional EQS-HBC users with $i$th person at a distance $d_i$ from the user under consideration, the signal to interference ratio (SIR) at that user's body will be given by:
\begin{equation}
    SIR = \frac{V_{Sig_{user}}}{V_{intf}}=\frac{V_{Sig_{user}}}{\sum_{i=1}^N V_{Sig_i}\times Coupling\left(d_i\right)}
    \label{eqn:interf}
\end{equation}
where $Coupling\left(d_i\right)$ is the inter-body coupling coefficient between the user under consideration, and interfering person.
Given the a signal level on desired user's body, $V_{Sig_{user}}$, equation \ref{eqn:interf} should be used to determine how many  other  EQS-HBC users, utilizing the same frequency band, could be tolerated in close proximity to the desired user.

\section*{Results}

For Radiative communication protocols such as Bluetooth, the signal decay between a transmitters and receiver is well-understood, and can be estimated well using the Friis Transmission Equation \cite{balanis_2016}:

\begin{equation}
    \frac{P_{Rx}}{P_{Tx}} \propto \left(\frac{\lambda}{d}\right)^2
\end{equation}
 
where $P_{Rx}$ and $P_{Tx}$ are received power and transmitted power respectively. This provides a simple outlook on the distance over which the signal from a radiative device can be picked up. For EQS-HBC devices however, a similar understanding is required in the electro-quasistatic region. We have motivated the fact that the coupling between two bodied in this regime is dictated by the coupling capacitance, $C_C$ between the two. In the following sections, we start with a biophysical model of EQS-HBC and extend that to incorporate this capacitive coupling between two human bodies to develop the first Bio-Physical Model for Inter-Body coupling in EQS regime. To understand the continuity from EQS to EM and the boundaries of EQS operation, we also discuss forms of coupling other than EQS over different frequency ranges going up to 1 GHz - where these devices behave as radiative devices instead. These theory and hypotheses are developed through rigorous analysis, simulation and measurements that we discuss in the following sections. Finally, utilizing this newfound understanding, we propose design strategies to minimize the security and interference risks of EQS inter-body coupling.




\subsection*{Different Frequency Regions of Inter-body Coupling}

\subsubsection*{Region 1 - Electro-quasistatic Coupling: Biophysical Modelling}
As mentioned before, compared to many commercial antenna designs, the human body has a large surface area. Naturally, this can introduce a capacitive coupling between two human subjects present close to each other. By modelling this coupling as a lumped capacitor, and using a simplified version of the capacitive HBC biophysical model developed by Maity et al \cite{bio_physical_tbme_19}, a basic circuit theoretic analysis can be performed. Two distinct cases are of interest, depending on the load impedance used at the receiver side - a low resistance load, typically 50 $\Omega$, and a capacitive load.
\begin{itemize}
    \item \textbf{ Resistive Load: ($Z_L=R_L$)} For many standard RF devices, use of a 50 $\Omega$ source and load impedance is the norm. This section explores the transfer characteristics assuming a pure capacitive coupling between two human bodies. The circuit model corresponding to this case can be found in fig. \ref{fig:concepts}d. The coupling capacitance $C_C$ and the load resistance $R_L$ together works as a high pass filter, and the pole of the filter depends on the exact coupling capacitance $C_C$ present between two human subjects, given a fixed load resistance $R_L$. This causes a 20 dB/decade rising slope in the channel gain vs frequency plot, until at higher frequencies - where the effect of a low pass filter formed by the source resistance $R_S$ and the body shunt-capacitance, $C_{Body}$ is encountered. The resulting response from circuit simulations can be seen in figure \ref{fig:concepts}e. 
    \item \textbf{Capacitive Load: ($Z_L=1/\omega C_L$)} For capacitive HBC, it has been suggested by Maity et al, that a capacitive load is a more viable option compared to a 50 $\Omega$ load, as it provides a flat-band frequency response in the low frequency. Now assuming the same receiver being present on a second subject, it should be interesting to see how much of the signal from the transmitting subject couples to the receiving subject. Simulating the circuit (fig \ref{fig:concepts}d) from this modality, shows a similar flat-band response in the low-frequency region, as shown in figure \ref{fig:concepts}e. A capacitive division is formed between the the coupling capacitance $C_C$ and the load capacitance $C_{eff, Rx}$, that division is independent of frequency, giving rise to the flat band. For frequencies above 100 MHz, a low pass effect is seen because of $R_S$ and $C_{tot}$.
\end{itemize}

Note that the plot shown in figure \ref{fig:concepts}f is result from a circuit simulation, assuming a lumped element model of figure \ref{fig:concepts}e. Of course, this modelling only holds in the EQS region (f < 1 MHz); the higher frequency regions will be explored in the following sections. In the EQS region, clearly a capacitive load ($Z_L=1/\omega C_L$) results into a consistently higher received voltage due to it's flat frequency response - as opposed to a 20 dB/decade rising slope for the resistive load ($Z_L=50\Omega$). Further, if a regular small antenna instead of a second human body is used as a coupler at the receiver (figure \ref{fig:concepts}b), the coupling capacitance $C_C$ would significantly drop - resulting into a much poor received voltage. In short, in the EQS region,
\begin{equation}
    V_{Rx,\ Antenna\ Coupled}\ll V_{Rx,\ Body\ Coupled,\ R_L} \ll V_{Rx,\ Body\ Coupled,\ C_L}
\end{equation}
So if an attacker is to device a strategy to snoop on an EQS-HBC device, the most effective strategy for them would be to use human body coupling, with an EQS-HBC receiver with a capacitive load.

\begin{figure}[!t]
    \centering
    \includegraphics[width=0.65\textwidth]{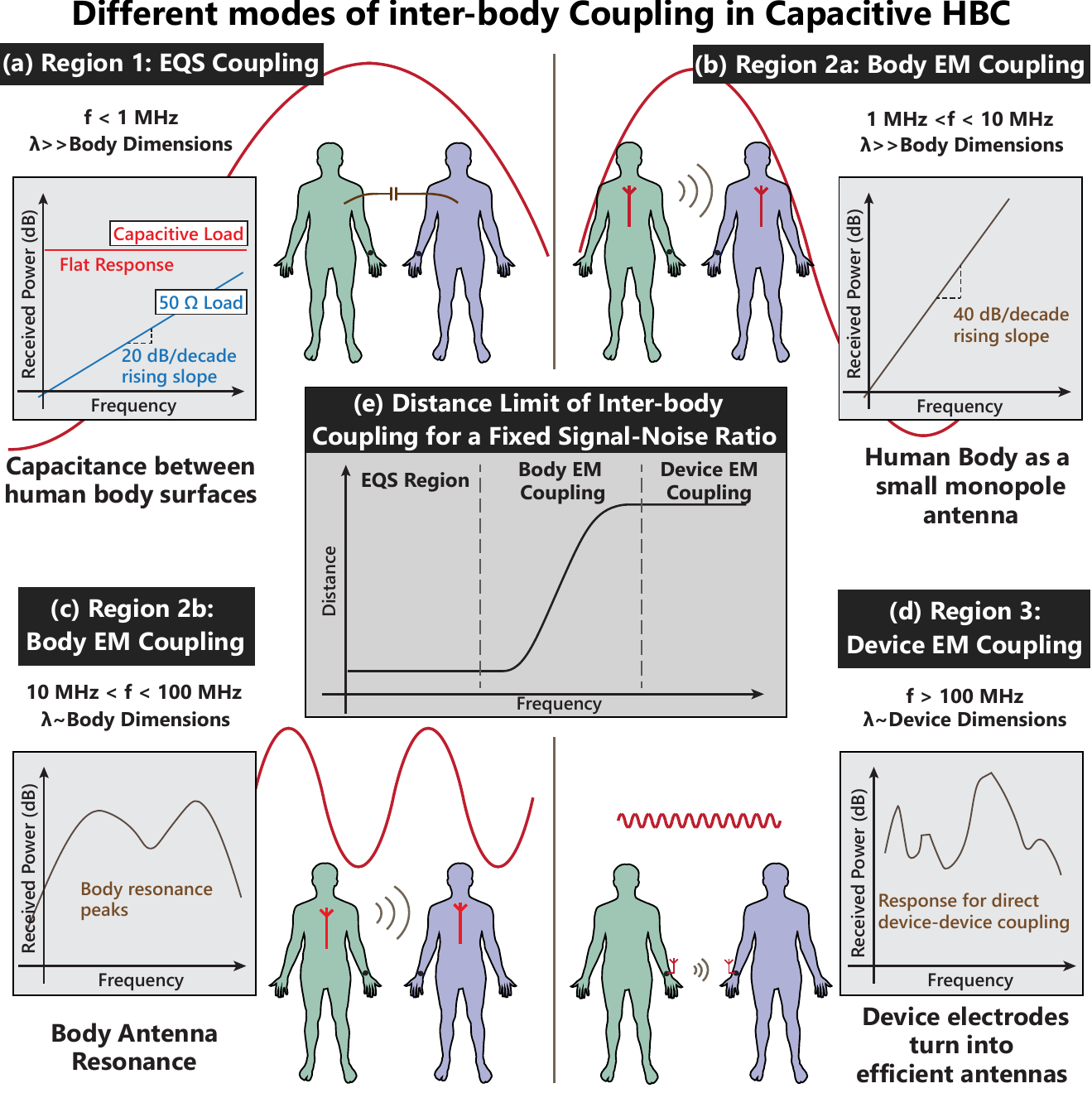}
    \caption{Inter-body coupling modes for Capacitive HBC users: (a) EQS Region, f<1 MHz, capacitive coupling dominates. (b) 1 MHz<f<10 MHz, body starts to act as a small monopole antenna, giving a 40dB/decade response in coupling. (c) 10 MHz<f<100 MHz, wavelength comparable to body dimensions; Body-antenna resonance peaks occur. (d) f>100 MHz, wavelength comparable to device dimensions; the devices start coupling through EM leakage. (e) The trend of maximum distance, over which signals from inter-body coupling can be detected (for a fixed SNR at the transmitter), over frequency. The distance limit is low independent of frequency for EQS coupling, increases rapidly once the two bodies start becoming efficient antennas, and becomes saturates once the devices become efficient antennas themselves. }
    \label{fig:coupling_modes}
\end{figure}

\subsubsection*{Region 2 - Inter-body Electromagnetic Coupling}
\begin{figure}[!t]
    \centering
    \includegraphics[width=0.7\textwidth]{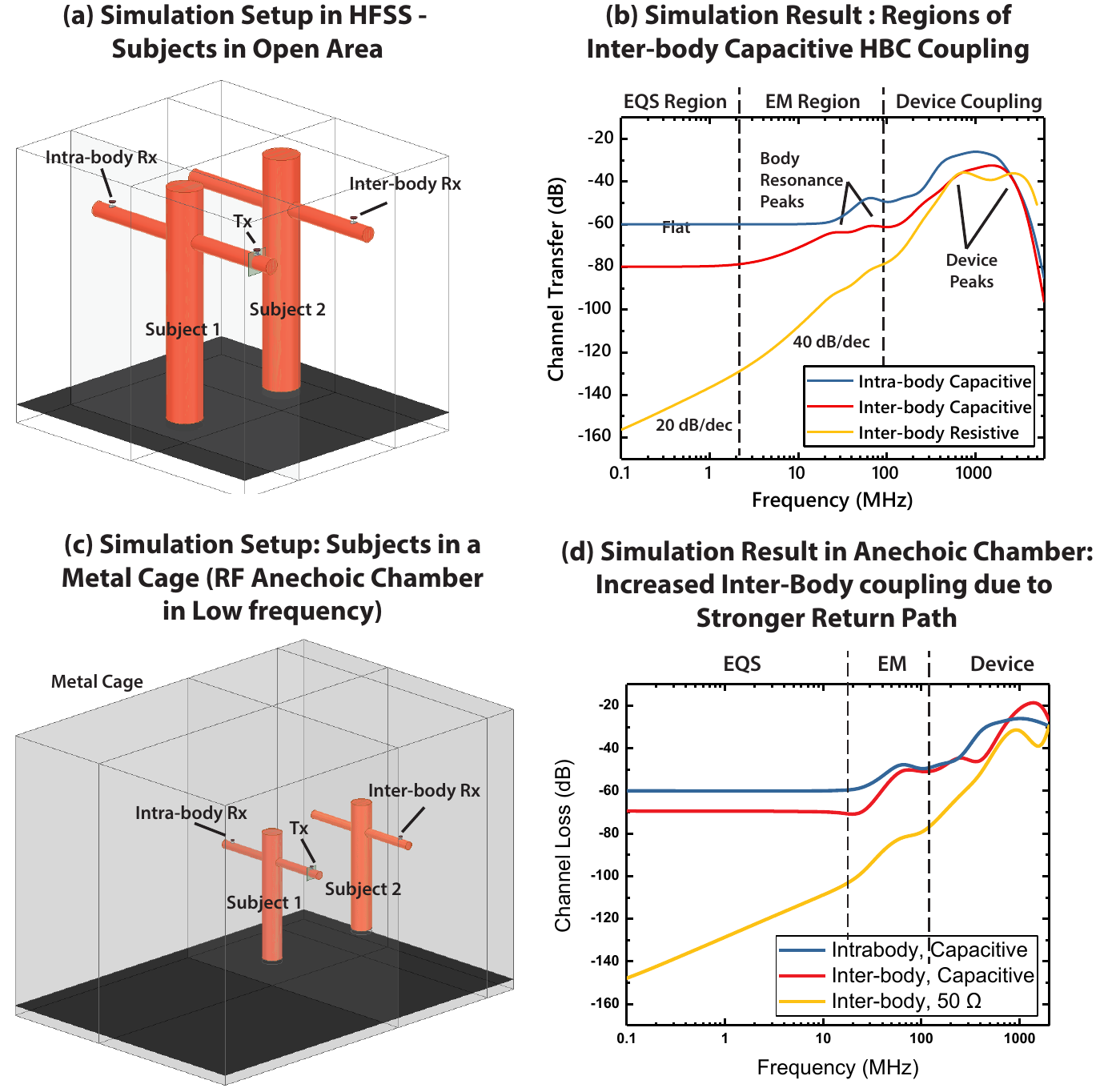}
    \caption{(a) Simulation setup used in ANSYS HFSS, using simplified models for the human body, and single-ended/ capacitive HBC electrodes as transmitter and receivers. Setup represents measurements in open-area. (b) Results from open-area simulation in HFSS. Three distinct regions are clearly visible, electro-quasistatic(EQS) region for freq$<$1 MHz, EM region for freq 1 MHz-100 MHz and device coupling region for freq$>$100 MHz. (c) Simulation setup for results in Anechoic Chamber. The subjects are enclosed in a metal-cage to represent higher return-path coupling in EQS region. (d) HFSS simulation results, inside anechoic chamber. The EQS region of the inter-body responses, are 10 dB higher compared to open air simulation results in \ref{fig:results}b. Because of this, the transition point between EQS region and EM region moves right.
    }
    
    \label{fig:results}
\end{figure}

Since the human body is made with conductive tissues, it is possible to look at a standing human subject as a cylinder, made with a weak conductor. As shown in fig \ref{fig:coupling_modes}b, a human subject standing on the earth's ground can be seen as a monopole antenna. To eliminate near field effects and only look at the electromagnetic antenna behaviour of the body, FEM simulations are performed in ANSYS HFSS, with plane-waves incident on the body and looking at the voltage induced on the body. For the purpose of simulation, as shown in figure \ref{fig:results}a, a simple model of the human body shaped as a crossed cylinders is used, with tissue properties of the muscle and skin. More details on the simulations can be found in the methods section. The results in figure \ref{fig:results}b show the response of the human body against electromagnetic waves, we will divide it into into two sections, as described below:

\begin{itemize}
    \item \textbf{Region 2a - Infinitesimal Monopole:} At low frequencies ($f<20 MHz$), wavelength $\lambda$ of the incident wave is large compared to the height $h$ of the subject. For that reason, the body can be thought of as an infinitesimal monopole at these frequencies. Now the radiation resistance of an infinitesimal monopole antenna of length $l$ is given by\cite{balanis_2016},
    
    \begin{equation}
        R_{rad}=80\pi^2(l/\lambda)^2
        \label{eqn:rad_res}
    \end{equation}
    
    so, the gain of the antenna is proportional to square of the frequency - $G_{Rx}\propto f^2$. So when the received power by the body is plotted in dB vs frequency, we should see a 20 dB/decade positive slope. As seen from the simulation results in fig \ref{fig:results}b, that is indeed the case. Note that when we look at the coupling between two human subjects later in the \textit{Combined Results} subsection, one of the human bodies act as a transmitting antenna, while the second as a receiving antenna. So, the net gain there is proportional to $f^4$, giving rise to a \textbf{40 dB/decade} slope in the gain vs frequency plot.
    
    \item \textbf{Region 2b - Body Resonance Peaks:} 
    For 10MHz<f<100MHz, the body dimensions become comparable to wavelength. As a result, antenna resonance peaks occur, as represented in Fig. \ref{fig:coupling_modes}c. The exact position and nature of the peaks will depend on the height and posture of the subjects.
    
\end{itemize}

\subsubsection*{Region 3 - Electromagnetic Coupling Between Devices/Electrodes} 
The electrodes of an HBC device that is used to couple HBC signal to a subject's body, are typically watch shaped, with a diameter of 3-5 cm. At frequencies $>$ 100 MHz, these electrodes start becoming efficient antennas themselves, peaking in the GHz range, depending on exact dimensions. As an example, if an electrode of diameter 5 cm is approximated as a mono-pole antenna of the same length, the resonant peak of the antenna occurs at 1.5 GHz in the air.

\subsection*{Results from FEM simulations and experiments}
We have discussed the different modalities through which signal transfer could happen between two human subjects, wearing an HBC transmitter and receiver respectively. In a real-world scenario, all these effects are present simultaneously, and depending on the region in the frequency spectrum one of these can become dominant. We show results from EM simulations as well as experiments in this section, to demonstrate this fact. For simulations, ANSYS HFSS, an FEM based Maxwell equation is solver is used. A simplified human body structure is assumed as shown in Fig \ref{fig:results}a. The details of the EM simulation setup can be found in the Methods section.

\begin{figure}[!t]
    \centering
    \includegraphics[width=0.7\textwidth]{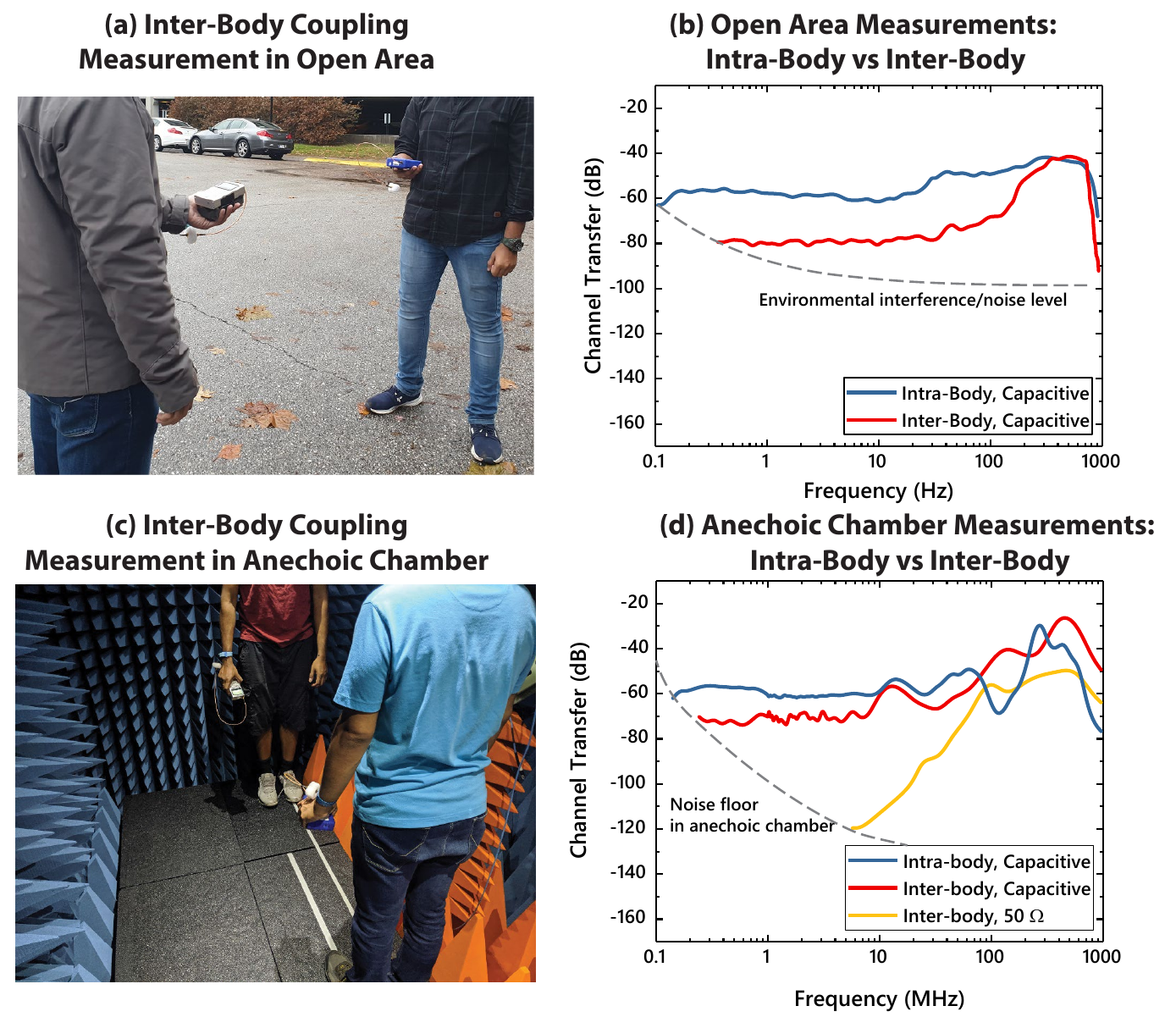}
    \caption{
    (a) Measurement setup in open-area. The subjects are kept at a distance of 1m for frequency sweep measurements. (b) Experiment results from open-area measurements.   Multiple transmitting devices are used to cover the whole frequency range, as shown later in fig \ref{fig:devices}a. (c) Measurement setup inside anechoic chamber. (d) Results from measurements inside the anechoic chamber.
    }
     \label{fig:expt_results}
\end{figure}

\subsubsection*{HFSS simulation for inter-body coupling transfer characteristics over frequency}
Two subjects are kept at a distance of 1m from each other, with capacitive HBC device models on stretched arms. Simulation is performed over the frequency range of 100 kHz - 1 GHz, for both capacitive and 50$\Omega$ termination at the receiving subject's device. The resulting transfer characteristics is shown in fig \ref{fig:results}b. As shown in fig \ref{fig:results}b, this transfer characteristics can broadly be divided into three regions, depending on the dominant modality of coupling in operation:

\begin{itemize}
    \item \textbf{Freq < 1 MHz:} In this region, we see a 20dB/decade rising slope for the 50 $\Omega$ termination, and a flat band response for the capacitive termination. This indicates that the dominant coupling method in this frequency range is electro-quasistatic, and hence can be modeled by circuit models shown in figure \ref{fig:concepts}c
    \item \textbf{Freq 1 MHz - 100 MHz:} In this region for the 50 $\Omega$ termination, we see a 40 dB/decade rising slope that flattens into peaks between 20 and 80 MHz. This indicates electromagnetic/ mono-pole antenna coupling between the two subjects. For capacitive termination, we see an increased response from the flat-band in the lower frequency range and peaks at similar frequencies as the 50 $\Omega$ termination. The slope is less than 40 dB/decade however, that indicates that both EQS and EM effects are equally present in this case, the EQS effect being a flat-band response at -80 dB, while the EM effect is a 40 dB/decade rising slope. When the two effects are added, a gentler rising slope results, and the peaks from the EM effect show up at a higher level (lower loss) compared to the 50 $\Omega$ termination.
    \item \textbf{Freq > 100 MHz:} In this frequency range, we see a sharp rise in the transfer characteristics, due to the electrodes becoming efficient antennas. This becomes the dominant mode of transfer, as in the GHz frequency range, the human body becomes an inefficient antenna. Its resonant frequency as a mono-pole antenna lies in the 20-80 MHz range, and the response starts dropping as the frequency is increased beyond that, as shown in figure \ref{fig:results}b. Transfer of signal by EQS capacitive coupling between the two subjects also becomes inefficient due to the previously discussed low-pass effect (Fig \ref{fig:concepts}e). So in the high frequency range ($>$100 MHz), a sniffing device that has a relatively small form-factor, like a hand-held antenna, can pick up leakage signal from an HBC device efficiently. However, as we'll discuss later, since a capacitive HBC device provides flat-band load, 
\end{itemize}

\subsubsection*{Experiments}
To validate the simulation results, we perform channel loss measurements between two human subjects inside an anechoic chamber. The subjects are kept at a distance of 1 meter, and a frequency sweep at the transmitter is performed from 100 kHz - 960 MHz. We use handheld devices for our measurements as opposed to wall connected devices, as wall connected devices share a common ground, and that reduces the channel loss, producing an optimistic result. Now to cover the entire frequency range of our experiments, we split the range into multiple handheld RF generators, details of which can be found in the Methods section. Note that the ground sizes of the different transmitting devices are slightly different, and that in turn makes the transmitter side return path capacitance, $C_{G,Tx}$ slightly different between the devices. As a result, there are discontinuities in the plot of the measurement data, in figure \ref{fig:results}f. 

While performing the experiments inside the anechoic chamber provides a controlled low-noise environment for gathering accurate frequency response, the chamber is enclosed a grounded metal cage and that affects the low frequency EQS region of the results. First, the grounded metal cage increases the overall return path capacitance, and that reduces channel loss. Second, as the EQS region now shows lower loss, the crossover point between the EQS and EM regions moves to a higher frequency. This is also supported by a second set of simulations where the anechoic chamber is also modelled. The result presented in fig \ref{fig:results}e shows a 10 dB reduction in loss in the EQS region compared to open-air simulation results in fig \ref{fig:results}b. Also, the cross-over point between EM and EQS region moves close to 10 MHz, as opposed to 1 MHz in fig \ref{fig:results}b.

From our examples in the anechoic chamber, we saw that intra-body and inter-body losses are 60 dB and 70 dB respectively. Since the anechoic chamber provides a strong return path, this is an optimistic estimate for inter-body channel loss. A pessimistic estimate of the channel loss comes from the open-air case, where the inter-body loss is about 80 dB. So even at 1 meter distance between the two bodies, the difference between intra-body and inter-body channel loss is between 10-20 dB.








\begin{figure}[!t]
    \centering
    \includegraphics[width=\textwidth]{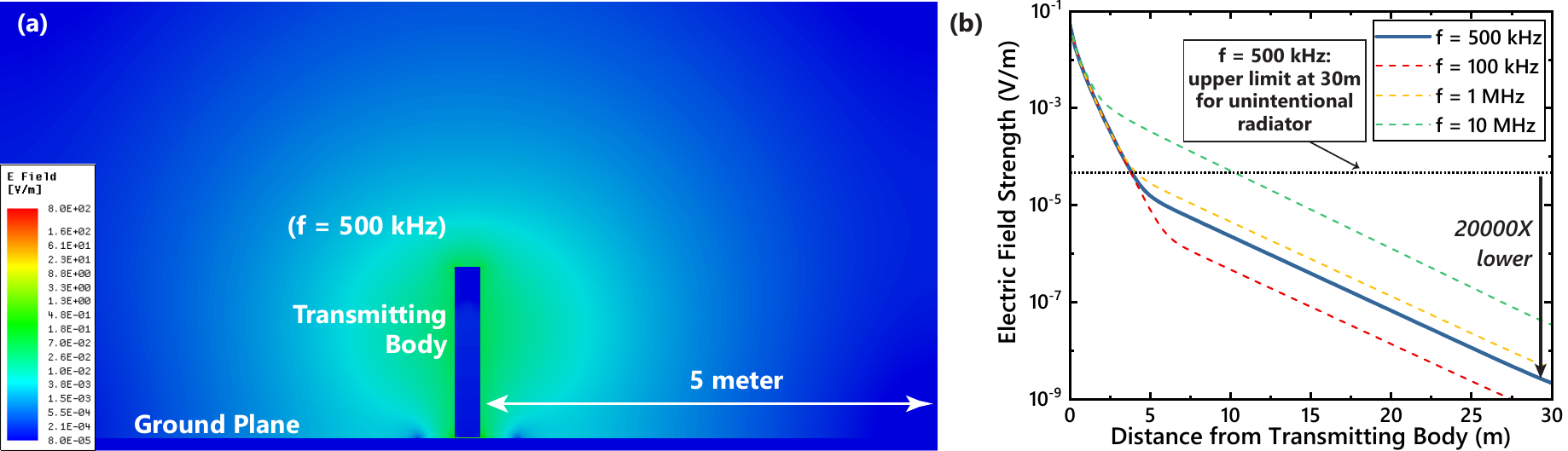}
    \caption{(a) Electric field decay from an EQS HBC device in a 2-D cross section at 500 kHz. (b) Plot of E-field decay vs distance shows that the E-Field drops 20000x below the FCC threshold to qualify as an unintentional radiator. This seconds the weak capacitive coupling demonstrated in Fig. \ref{fig:concepts}(b)}
    \label{fig:fcc}
\end{figure}

\subsection*{FCC Regulations: Can EQS-HBC Device be classified as an Unintentional Radiator?}

In the previous section we have looked into the inter-body coupling among humans when they are using EQS-HBC as BAN communication. This coupling depends on the Electric fields created by the EQS HBC User and the surface area of the recipient. Related to the phenomena of electric fields around the human body during EQS HBC transmission, an important question arises about the usability of these devices in practice:  Can EQS-HBC Device be classified as an Unintentional Radiator?

According to FCC regulations\cite{fcc_regulations} as shown in table \ref{tab:fcc}, the definition of intentional vs unintentional radiator is as follows: for a frequency F between 9-490 kHz, if the fields at 300m distance are below 2400/F and for a for a frequency F between 490 kHz-1.705 MHz, if the fields at 30m distance are below 24000/F, the device can be classified as an unintentional radiator - which means no additional FCC certification would be required for deployment of these devices in practice. Using our analysis and theory that we have developed, we can get a great sense of the radiated filed. In figure \ref{fig:fcc}a, electric field emission from a human body using an EQS-HBC transmitter is visualized. From the decay of the field vs distance plotted in figure \ref{fig:fcc}b, it can be seen that at 30m distance, the electric field is about 20000 times lower than the required FCC limit (table \ref{tab:fcc}). So, the fields emanated from the EQS-HBC devices so low - that it is not perceptible by other devices and hence EQS-HBC devices can be classified as unintentional radiators, and can be deployed without the need for new standards and certifications.

\begin{table}[!t]
\centering
\begin{tabular}{@{}ccc@{}}
\toprule
Frequency (MHz) & Field Strength ($\mu$V/m) & Measurement Distance (m) \\ \midrule
0.009-0.490     & 2400/F (kHz)           & 300                      \\
0.490-1.705     & 24000/F (kHz)          & 30                       \\
1.705-30.0      & 30                    & 30                       \\
30-88           & 100                   & 3                        \\
88-216          & 150                   & 3                        \\
216-960         & 200                   & 3                        \\
Above 960       & 500                   & 3                        \\ \bottomrule
\end{tabular}
\caption{FCC Field Limit Regulations for Unintentional Radiators \cite{fcc_regulations}}
\label{tab:fcc}
\end{table}

\subsection*{EQS-HBC Design Consideration to Maximally Protect against Inter-Body Attack and Interference}

\begin{figure}
    \centering
    \includegraphics[width=\textwidth]{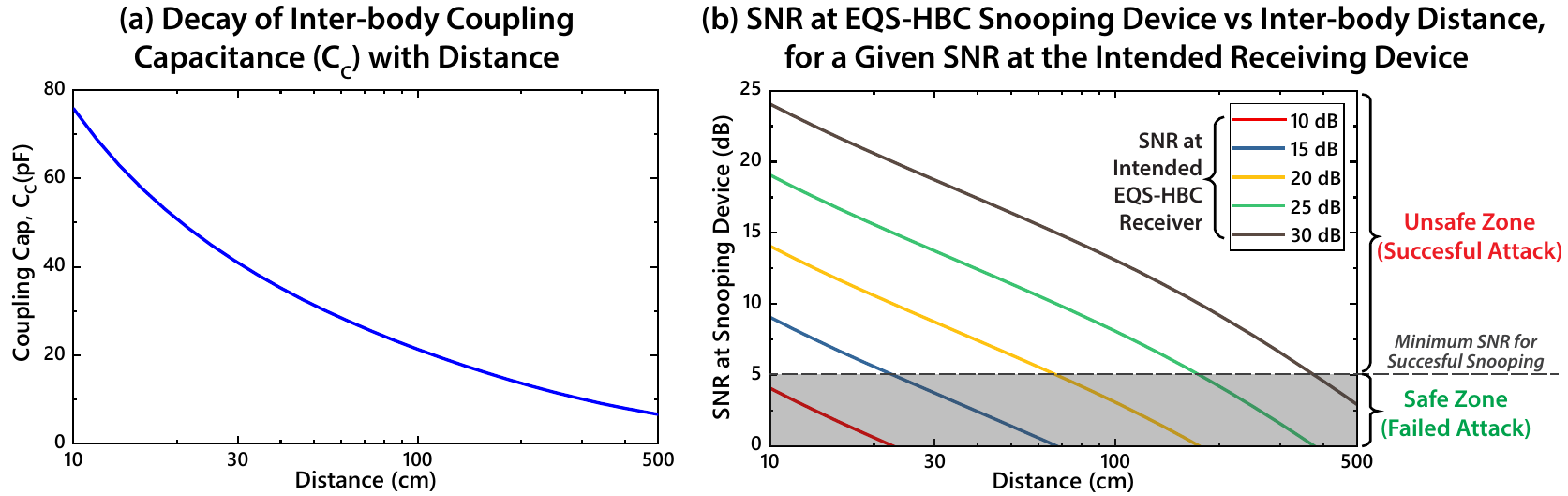}
    \caption{(a) Simulated inter-body coupling capacitance ($C_C$) with distance. (b) SNR at a snooping device, for a given SNR at the intended receiving device. When the SNR at the snooping device falls below 6-9 dB, successful attack is prevented.}
    \label{fig:snr}
\end{figure}

We have shown the different regimes of inter-body coupling in HBC. More specifically, for EQS-HBC (f$<$1 MHz), inter- body transfer characteristics show a flat band response similar to intra-body HBC, given a capacitive load is used at the receiving device. The difference in the channel loss between these two cases determines whether a successful attack can be performed using the human body as a capacitive coupler. By comparing the equations shown in figure \ref{fig:concepts}d and \ref{fig:concepts}e, the difference in channel loss between these two cases can be given by:
\begin{equation}
    \frac{V_{Inter-body}(d)}{V_{Intra-body}}=\frac{C_C(d)}{C_{Body}}
    \label{eqn:extra_loss}
\end{equation}
where $d$ is the distance between the EQS-HBC user and the attacker's body. Since the body to ground capacitance $C_{Body}$ is fixed at around 150 pF\cite{bio_physical_tbme_19}, variation of the inter-body coupling capacitance $C_C$ with $d$ will determine the variation of $V_{Inter-body}$ with $d$. Figure \ref{fig:snr}(a) shows a plot of $C_C$ vs $d$, obtained by electrostatic FEM simulation in ANSYS Maxwell. Accordingly, putting $C_C=21.pF$ at $d=1m$ in equation \ref{eqn:extra_loss}, we find out an additional loss of ~17 dB for inter-body coupling. This matches with our previous experimental finding, where we saw that at 1 meter distance, the difference between intra-body and inter-body channel loss was about 10-20 dB. At 5m, $C_C$ reduces to 6.6 pF, which raises this difference to 27 dB.

Now, let us consider the EQS-HBC using On-Off Keying (OOK) signals\cite{maity_wearable_2017,maity2019bodywire}. Since OOK requires a signal to noise ratio (SNR) of at least 6dB to ensure proper operation, transmitter power should usually be kept 6-9dB greater than the receiver sensitivity including channel loss. Using equation \ref{eqn:extra_loss} and $C_C$ from figure \ref{fig:snr}(a), we can estimate the SNR at the snooping person's receiver for a pre-determined SNR at the intended receiver on the EQS-HBC user's body. This is plotted in figure \ref{fig:snr}b for set of given SNR at the intended receiver. By staying in the shaded region in the plot, a successful attack can be prevented. For example, if the signal level of the EQS-HBC transmitter is set to maintain an SNR of 10 dB at the intended receiver, an attacker will not be able to snoop that signal even at 10 cm distance from the user. Further, by setting the signal level in this way, interference effects are also reduced between multiple adjacent EQS-HBC users in a common space. So even if inter-body coupling in EQS-HBC introduces a risk of unintended signal sniffing and/or interference, steps can be taken towards setting the signal level of an HBC device to minimize or eliminate the possibility of the same.

\section*{Conclusion}

In conclusion, we show for the first time that the human body can function a capacitive coupler to pick up EQS-HBC signals, making this BAN technique vulnerable to inter-body attack and interference. We explore inter-body coupling modalities over a broad frequency range (100 KHz - 1 GHz), identify and explain three distinct regions - namely EQS inter-body coupling, inter-body EM coupling and inter-device EM coupling. We develop the first biophysical model that describes inter-body coupling in the EQS frequency region (< 1 MHz) as a function of the capacitance between two human bodies, which in turn is a function of distance. Finally, we demonstrate that by optimizing the signal level at a EQS-HBC transmitting device, the inter-body coupling vulnerabilities can be reduced (if not eliminated) to a distance of less than 10 cm of an EQS-HBC user's body, restoring the physical security  of EQS-HBC.

\section*{Methods}
This section contains details regarding our simulation and experimental methods, to facilitate reproduction of the results if anyone wishes to do so.
\subsection*{EM Simulation Setup}
All the EM simulations have been performed in Ansoft HFSS, which is an Finite Element Methods based Maxwell Equation solver. Two different types of models were used; a simple cylindrical model for developing intuition, and a detailed model consisting of different human tissue parts, for more accurate results. Dielectric properties of all body tissues have been taken from the works of Gabriel et al \cite{Gabriel_measurements_1996}.
\subsubsection*{Simple Model} A simple model was created using two perpendicular cylinders, as shown in figure \ref{fig:results}a, representing the torso and extended arms. The radius of the cylinders were 14 cm and 6 cm respectively. The height of torso was taken to be 180 cm, and the entire arm span was taken to be 180 cm as well. Both the torso and the arms were divided into a 4 mm outer shell of skin, and an interior of muscle. This crossed-cylinder model were floated 2 cm above a plane with Perfect E Boundary in HFSS, supposed to replicated an infinite ground plane. A rubber cylinder of same diameter as the torso was placed between the torso and the perfect E plane. The entire model was then enclosed in a region of air, measuring 120 cm x 60 cm x 340 cm. Excitation for the simulation was provided through capacitive coupling, as described in the next sub-section.
\subsubsection*{Excitation} A capacitive coupling model is used to provide excitation to the body attached to a transmitter. The coupler consists of two copper discs with a radius of 2.5 cm. One of the discs, was 2 mm thick and was curved onto the arm - this disc replicated an electrode patch attached to the arm. The other disc with a thickness of 5 mm, replicated the ground plane of an wearable watch-like HBC device. the separation between the two plates can be varied to change the capacitance between the plates, a distance of 3 cm was used in our simulations, yielding an approximate parallel plate capacitance of 0.6 pF. Alternatively, a fixed capacitance of choice can be maintained between the plates, using a lumped RLC boundary in HFSS. A voltage source excitation was placed between the two plates. In HFSS, this imparts an alternating potential difference of 1 V p-p between the two plates, replicating an ideal AC voltage source. This is unlike the lumped port excitation method, which is ideal for $50\Omega$ matched excitations, but may give rise to unexpected reflections when coupling to a non-standard RF model. 

\subsubsection*{Measuring of voltage at receiver} The receiving node structure is identical to that of the transmitter. A lumped RLC boundary is placed between the electrode and the ground plates at the receiver, which was set to $50\Omega$ for a low impedance termination, and 1 pF for high impedance termination cases. The potential difference between the plates was calculated by integrating the electric field along a straight line between the electrode and ground plates.

\subsubsection*{Detailed Model}
  The basic simulation setup is now carried over to a detailed model of human body, to find out values of E and H fields that different body parts would experience for a certain HBC operating voltage. The human body model used for all the simulations was obtained from NEVA Electromagnetics LLC. The specific model used is the VHP-Female v2.2 \cite{neva_model}, which was generated from a 162 cm tall, 60 year old female subject. Similar to the simulations of the cylindrical dummy, the excitation were provided by a single disc and a floating ground plate in Capacitive HBC, and two spaced disc in case of galvanic HBC. The dielectric properties of the body tissues were adapted from the works of Gabriel et al.\cite{Gabriel_measurements_1996}; we did not use the material properties that came packaged with the HFSS version of the NEVA EM model, as the HFSS model did not incorporate tissue properties for frequencies less than 10 MHz.

\subsection*{Experimental Setup}
Experiments are conducted in two parts - the first set of measurements are made inside an EM anechoic chamber to maintain a controlled environment and achieve noise immunity. The second sets of experiment are done in an open area such as an empty parking lot, to compare signal levels with the ones inside anechoic chamber. For the purpose of replicating real-world HBC devices, hand-held transmitting and receiving devices were used, as opposed to wall connected such as a Vector Network Analyzer. Wall connected devices essentially share a common ground and bring the ground planes of the transmitter and receiver at a common potential, thus showing a lower loss and giving an optimistic channel transfer characteristics\cite{nath_return_cap,bio_physical_tbme_19}. 

\begin{figure}
    \centering
    \includegraphics[width=\textwidth]{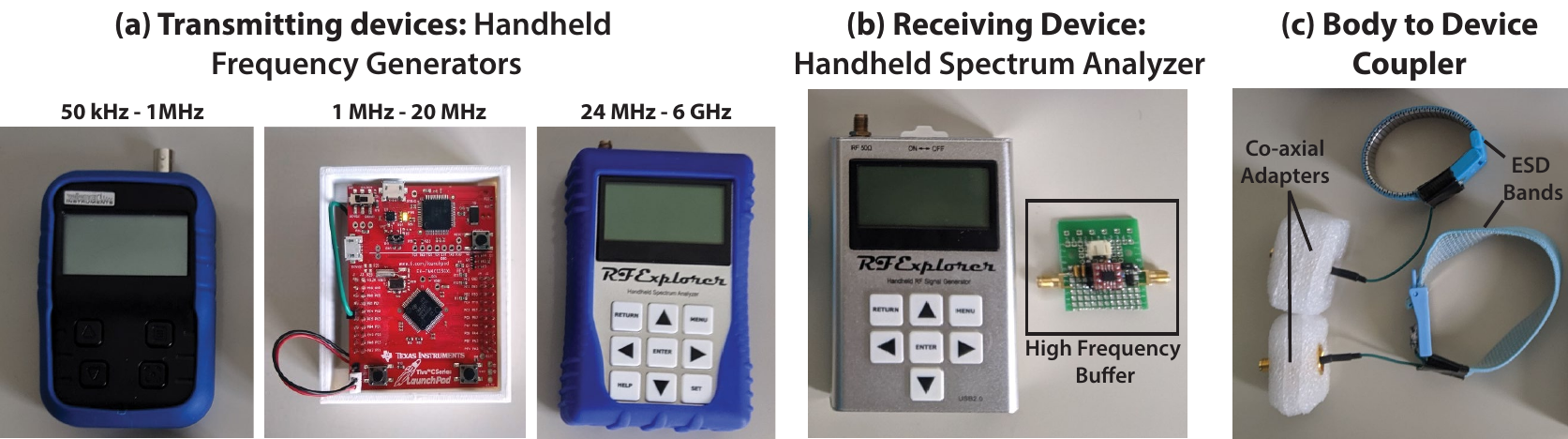}
    \caption{Devices used for experiments: (a) Transmitting devices: Velleman Handheld RF Generator for freq$<$1 MHz, signal generator built using a Tiva C Launchpad Board for freq 1MHz-20MHz and RF Explorer Handheld RF generator for freq$>$24 MHz. (b) RF Explorer Handheld spectrum analyzer used as a receiving device, and a high-frequency buffer used for high impedance/capacitive load measurements. (c) Couplers used to connect the transmitter and receiver devices to body.}
    \label{fig:devices}
\end{figure}

\subsubsection*{Transmitting Devices}
We plot the transfer characteristics over a large frequency range, namely $~$ 100kHz-1GHz. We use multiple handheld RF signal generators to cover the entire range:
\begin{itemize}
    \item \textbf{Freq $<$ 1 MHz} A hand-held signal generator from Velleman is used to generate sinusoidal signal for frequencies lower than 1 MHz.
    \item \textbf{Freq 1 MHz - 20 MHz} An in-house signal generator is used, built using a Texas Instruments Tiva C Launchpad evaluation board. The generator provides a square wave, the fundamental harmonic is used for our experiments.
    \item \textbf{Freq $>$ 24 MHz } RF explorer handheld RF generator is used, which generates sine wave in the range 24 MHz - 6 GHz
\end{itemize}

All the transmitting devices are characterized using a precision spectrum analyzer, to record accurate transmitting power of the fundamental at each frequency point.

\subsubsection*{Receiving Devices}
We used a handheld spectrum analyzer from RF Explorer that covers 50 kHz - 960 MHz. The range was adjusted for each frequency point measurement, to include only the fundamental peak, and the peak power was noted. Subtracting the characterized transmitter power from this received power provides the channel transfer gain at that frequency. For measuring the 50 $\Omega$ termination cases, the spectrum analyzer is directly connected to the HBC coupler, as the device has an input impedance of 50 $\Omega$. For capacitive or high impedance load measurements, a high-frequency buffer is connected to the HBC coupler first, and the buffer's output is given to the spectrum analyzer. The buffer board, shown in fig \ref{fig:devices}b, is made using BUF602, a high speed buffer from Texas Instruments. The board is configured to have an input resistance of 1 M$\Omega$, so that the input impedance essentially becomes capacitive in the frequency of operation.

\subsubsection*{Body to Device Coupler} To couple the transmitting and receiving devices to a subject's body, an ESD wristband is used, worn on the subject's arm. The signal pin of a co-ax adapter is connected to the metal plate of the ESD band by soldering a small piece of wire. An example of the coupler is shown in fig \ref{fig:devices}c. These couplers are in turn connected to the transmitting and receiving devices using a shielded co-ax cable. 

\subsubsection*{Frequency Sweep Measurements}

To obtain leakage or inter-body transfer characteristics over frequency, the experiment subjects are asked to stand 1 meter apart, facing each other. The transmitting device is coupled to one subject while the receiving device to the other. The subjects operate the handheld transmitting and receiving devices themselves to sweep frequency by hand. The receiving subject communicates the resulting peak power to a third person standing away from the two subjects, to log data.


\bibliography{body_antenna}

\begin{thebibliography}{10}
\urlstyle{rm}
\expandafter\ifx\csname url\endcsname\relax
  \def\url#1{\texttt{#1}}\fi
\expandafter\ifx\csname urlprefix\endcsname\relax\def\urlprefix{URL }\fi
\expandafter\ifx\csname doiprefix\endcsname\relax\def\doiprefix{DOI: }\fi
\providecommand{\bibinfo}[2]{#2}
\providecommand{\eprint}[2][]{\url{#2}}

\bibitem{sen_tedx2019}
\bibinfo{author}{Sen, S.}
\newblock \bibinfo{title}{{How your body can play an integral role in wearable
  security}{ | }{TedX Indianapolis}}.
\newblock
  \bibinfo{howpublished}{\url{https://www.ted.com/talks/shreyas_sen_how_your_body_will_play_an_integral_role_in_the_future_of_wearable_security}}
  (\bibinfo{year}{2019}).
\newblock \bibinfo{note}{[accessed March 5, 2020]}.

\bibitem{nsr_das_19}
\bibinfo{author}{Das, D.}, \bibinfo{author}{Maity, S.},
  \bibinfo{author}{Chatterjee, B.} \& \bibinfo{author}{Sen, S.}
\newblock \bibinfo{journal}{\bibinfo{title}{Enabling covert body area network
  using electro-quasistatic human body communication}}.
\newblock {\emph{\JournalTitle{Scientific Reports}}}
  \textbf{\bibinfo{volume}{9}}, \bibinfo{pages}{4160--2906},
  \doiprefix\url{https://doi.org/10.1038/s41598-018-38303-x}
  (\bibinfo{year}{2019}).

\bibitem{bodywire_jssc_19}
\bibinfo{author}{{Maity}, S.}, \bibinfo{author}{{Chatterjee}, B.},
  \bibinfo{author}{{Chang}, G.} \& \bibinfo{author}{{Sen}, S.}
\newblock \bibinfo{journal}{\bibinfo{title}{Bodywire: A 6.3-pj/b 30-mb/s
  −30-db sir-tolerant broadband interference-robust human body communication
  transceiver using time domain interference rejection}}.
\newblock {\emph{\JournalTitle{IEEE Journal of Solid-State Circuits}}}
  \textbf{\bibinfo{volume}{54}}, \bibinfo{pages}{2892--2906}
  (\bibinfo{year}{2019}).

\bibitem{tomovski_sar_2011}
\bibinfo{author}{Tomovski, B.}, \bibinfo{author}{Gräbner, F.},
  \bibinfo{author}{Hungsberg, A.}, \bibinfo{author}{Kallmeyer, C.} \&
  \bibinfo{author}{Linsel, M.}
\newblock \bibinfo{journal}{\bibinfo{title}{Effects of electromagnetic field
  over a human body, sar simulation with and without nanotextile in the
  frequency range 0.9-1.8ghz}}.
\newblock {\emph{\JournalTitle{Journal of Electrical Engineering}}}
  \textbf{\bibinfo{volume}{62}}, \bibinfo{pages}{349--354},
  \doiprefix\url{10.2478/v10187-011-0055-6} (\bibinfo{year}{2011}).

\bibitem{kibret_sar_2014}
\bibinfo{author}{Kibret, B.}, \bibinfo{author}{Teshome, A.} \&
  \bibinfo{author}{Lai, D.}
\newblock \bibinfo{journal}{\bibinfo{title}{Analysis of the whole-body averaged
  specific absorption rate (sar) for far-field exposure of an isolated human
  body using cylindrical antenna theory}}.
\newblock {\emph{\JournalTitle{Progress In Electromagnetics Research M}}}
  \textbf{\bibinfo{volume}{38}}, \bibinfo{pages}{103--112},
  \doiprefix\url{10.2528/PIERM14072201} (\bibinfo{year}{2014}).

\bibitem{kibret_cyl_2015}
\bibinfo{author}{Kibret, B.}, \bibinfo{author}{Teshome, A.} \&
  \bibinfo{author}{Lai, D.}
\newblock \bibinfo{journal}{\bibinfo{title}{Cylindrical antenna theory for the
  analysis of whole-body averaged specific absorption rate}}.
\newblock {\emph{\JournalTitle{IEEE Transactions on Antennas and Propagation}}}
  \textbf{\bibinfo{volume}{63}}, \bibinfo{pages}{5224 -- 5229},
  \doiprefix\url{10.1109/TAP.2015.2478484} (\bibinfo{year}{2015}).

\bibitem{kibret_antenna_hbc_2014}
\bibinfo{author}{Kibret, B.}, \bibinfo{author}{Teshome, A.} \&
  \bibinfo{author}{Lai, D.}
\newblock \bibinfo{journal}{\bibinfo{title}{Human body as antenna and its
  effect on human body communications}}.
\newblock {\emph{\JournalTitle{Progress In Electromagnetics Research}}}
  \textbf{\bibinfo{volume}{148}}, \bibinfo{pages}{193--207},
  \doiprefix\url{10.2528/PIER14061207} (\bibinfo{year}{2014}).

\bibitem{Hwang_interf_2017}
\bibinfo{author}{{Hwang}, J.}, \bibinfo{author}{{Kang}, T.},
  \bibinfo{author}{{Kwon}, J.} \& \bibinfo{author}{{Park}, S.}
\newblock \bibinfo{journal}{\bibinfo{title}{Effect of electromagnetic
  interference on human body communication}}.
\newblock {\emph{\JournalTitle{IEEE Transactions on Electromagnetic
  Compatibility}}} \textbf{\bibinfo{volume}{59}}, \bibinfo{pages}{48--57},
  \doiprefix\url{10.1109/TEMC.2016.2598582} (\bibinfo{year}{2017}).

\bibitem{kibret_monopole_2015}
\bibinfo{author}{Kibret, B.}, \bibinfo{author}{Teshome, A.} \&
  \bibinfo{author}{Lai, D.}
\newblock \bibinfo{journal}{\bibinfo{title}{Characterizing the human body as a
  monopole antenna}}.
\newblock {\emph{\JournalTitle{IEEE Transactions on Antennas and Propagation}}}
  \textbf{\bibinfo{volume}{63}}, \bibinfo{pages}{4384 -- 4392},
  \doiprefix\url{10.1109/TAP.2015.2456955} (\bibinfo{year}{2015}).

\bibitem{Li_antenna_2017}
\bibinfo{author}{Li, J.}, \bibinfo{author}{Nie, Z.}, \bibinfo{author}{Liu, Y.},
  \bibinfo{author}{Wang, L.} \& \bibinfo{author}{Hao, Y.}
\newblock \bibinfo{journal}{\bibinfo{title}{Evaluation of propagation
  characteristics using the human body as an antenna}}.
\newblock {\emph{\JournalTitle{Sensors (Basel, Switzerland)}}}
  \textbf{\bibinfo{volume}{17}}, \bibinfo{pages}{2878},
  \doiprefix\url{10.3390/s17122878} (\bibinfo{year}{2017}).

\bibitem{makehuman}
\bibinfo{title}{Makehuman: Open source tool for making 3d characters}.
\newblock \bibinfo{howpublished}{\url{http://www.makehumancommunity.org/}}.
\newblock \bibinfo{note}{[accessed March 5, 2020]}.

\bibitem{maity_forward_loss_2018}
\bibinfo{author}{Maity}, \bibinfo{author}{Mojabe} \& \bibinfo{author}{Sen}.
\newblock \bibinfo{journal}{\bibinfo{title}{Characterization of human body
  forward path loss and variability effects in voltage-mode hbc}}.
\newblock {\emph{\JournalTitle{IEEE Microwave and Wireless Components
  Letters}}} \textbf{\bibinfo{volume}{28}}, \bibinfo{pages}{266--268},
  \doiprefix\url{10.1109/LMWC.2018.2800529} (\bibinfo{year}{2018}).

\bibitem{nath_return_cap}
\bibinfo{author}{{Nath}, M.}, \bibinfo{author}{{Maity}, S.} \&
  \bibinfo{author}{{Sen}, S.}
\newblock \bibinfo{journal}{\bibinfo{title}{Towards understanding the return
  path capacitance in capacitive human body communication}}.
\newblock {\emph{\JournalTitle{IEEE Transactions on Circuits and Systems II:
  Express Briefs}}} \bibinfo{pages}{1--1},
  \doiprefix\url{10.1109/TCSII.2019.2953682} (\bibinfo{year}{2019}).

\bibitem{bio_physical_tbme_19}
\bibinfo{author}{{Maity}, S.} \emph{et~al.}
\newblock \bibinfo{journal}{\bibinfo{title}{Bio-physical modeling,
  characterization, and optimization of electro-quasistatic human body
  communication}}.
\newblock {\emph{\JournalTitle{IEEE Transactions on Biomedical Engineering}}}
  \textbf{\bibinfo{volume}{66}}, \bibinfo{pages}{1791--1802},
  \doiprefix\url{10.1109/TBME.2018.2879462} (\bibinfo{year}{2019}).

\bibitem{balanis_2016}
\bibinfo{author}{Balanis, C.~A.}
\newblock \emph{\bibinfo{title}{Antenna Theory: Analysis and Design}}
  (\bibinfo{publisher}{Wiley Blackwell}, \bibinfo{year}{2016}),
  \bibinfo{edition}{4} edn.

\bibitem{fcc_regulations}
\bibinfo{title}{Electronic code of federal regulations e-cfr title 47 part 15,
  subpart c, 15.209}.
\newblock
  \bibinfo{howpublished}{\url{https://ecfr.io/Title-47/se47.1.15_1209}}.
\newblock \bibinfo{note}{[accessed March 5, 2020]}.

\bibitem{maity_wearable_2017}
\bibinfo{author}{Maity, S.}, \bibinfo{author}{Das, D.} \& \bibinfo{author}{Sen,
  S.}
\newblock \bibinfo{title}{Wearable health monitoring using capacitive
  voltage-mode {Human} {Body} {Communication}}.
\newblock In \emph{\bibinfo{booktitle}{2017 39th {Annual} {International}
  {Conference} of the {IEEE} {Engineering} in {Medicine} and {Biology}
  {Society} ({EMBC})}}, \bibinfo{pages}{1--4},
  \doiprefix\url{10.1109/EMBC.2017.8036748} (\bibinfo{year}{2017}).

\bibitem{maity2019bodywire}
\bibinfo{author}{Maity, S.} \emph{et~al.}
\newblock \bibinfo{journal}{\bibinfo{title}{Bodywire: A 6.3-pj/b 30-mb/s- 30-db
  sir-tolerant broadband interference-robust human body communication
  transceiver using time domain interference rejection}}.
\newblock {\emph{\JournalTitle{IEEE Journal of Solid-State Circuits}}}
  \textbf{\bibinfo{volume}{54}}, \bibinfo{pages}{2892--2906}
  (\bibinfo{year}{2019}).

\bibitem{Gabriel_measurements_1996}
\bibinfo{author}{Gabriel, S.}, \bibinfo{author}{Lau, R.~W.} \&
  \bibinfo{author}{Gabriel, C.}
\newblock \bibinfo{journal}{\bibinfo{title}{The dielectric properties of
  biological tissues: {II}. measurements in the frequency range 10 hz to 20
  {GHz}}}.
\newblock {\emph{\JournalTitle{Physics in Medicine and Biology}}}
  \textbf{\bibinfo{volume}{41}}, \bibinfo{pages}{2251--2269},
  \doiprefix\url{10.1088/0031-9155/41/11/002} (\bibinfo{year}{1996}).

\bibitem{neva_model}
\bibinfo{title}{{NEVA Electromagnetics LLC}{ | }{Static VHP-Female model v2.2 -
  VHP-Female College}}.
\newblock
  \bibinfo{howpublished}{\url{https://www.nevaelectromagnetics.com/vhp-female-2-2}}.
\newblock \bibinfo{note}{[accessed March 5, 2020]}.

\end{thebibliography}

\section*{Acknowledgements}

This work was supported by Eli Lilly and Company through the Connected Health care initiative and in-part by the Air Force Office of Scientific Research (AFOSR) Young Investigator Award under Grant No. FA9550-17-1-0450. The authors would like to thank PhD students Debayan Das, David Yang, Donghyun Seo, Baibhab Chatterjee, Nirmoy Modak, Arunashish Datta and Faizul M. Bari; as well as Visiting Scholar Gargi Bhattacharya at Purdue University for their immense co-operation and support during the experiments.

\section*{Author contributions statement}

M.N., S.M., S.W. and S.S. conceived the idea. S.A. designed the in-house EQS-HBC transmitter for experiments. M.N. conducted the theoretical analysis, numerical simulations, and performed the experiments with help from S.M. and supervision from S.W. and S.S. All authors contributed to the drafting of this manuscript, and have read and approved the final version of the manuscript.


\section*{Additional information}

\textbf{Competing interests} The authors declare no competing interests.


%



\end{document}